\documentstyle[aps,prl,epsfig]{revtex}

\newcommand{\rH}{r_{\scriptscriptstyle\mathrm H}}

\begin{document}

\draft

\wideabs{
  \title{Entropy of constant curvature black holes in general relativity}
  \author{Jolien D. E. Creighton}
  \address{Theoretical Astrophysics, California Institute of Technology,
    Pasadena CA 91125}
  \author{Robert B. Mann}
  \address{Department of Physics, University of Waterloo, Waterloo, Ontario,
    Canada N2L~3G1}
  \date{\today}
  \maketitle
  \begin{abstract}%
    We consider the thermodynamic properties of the constant curvature black
    hole solution recently found by Ba{\~n}ados.  We show that it is possible
    to compute the entropy and the quasilocal thermodynamics of the spacetime
    using the Einstein-Hilbert action of General Relativity.  The constant
    curvature black hole has some unusual properties which have not been seen
    in other black hole spacetimes.  The entropy of the black hole is not
    associated with the event horizon; rather it is associated with the region
    between the event horizon and the observer.  Further, surfaces of constant
    internal energy are not isotherms so the first law of thermodynamics exists
    only in an integral form. These properties
    arise from the unusual topology of the Euclidean black hole instanton.
  \end{abstract}
}

\narrowtext

It is generally believed that black holes have a fundamental role to play
in furthering our understanding of the quantization of gravity.  Indeed,
a wide variety of spacetimes representing black holes with unusual
properties has been discoverd in the past decade as a consequence of an
intensive study of the various approaches to quantum gravity. Further
progress will necessarily entail a more thorough investigation of
the basic thermodynamics of the different species of black holes.

A new type of black hole solution
has been found recently by Ba{\~n}ados~\cite{b:1997}.  This solution, 
which is one possible generalization of the 2+1 dimensional black
hole~\cite{bhtz:1993} to higher dimensions, represents a black hole in a
spacetime with toroidal topology and constant curvature.  The constant
curvature black hole (CCBH) is essentially anti-de\thinspace Sitter spacetime
with identifications, and so it is a solution of any theory which contains
anti-de\thinspace Sitter spacetime.%
\footnote{An examination of all identifications
  in four dimensional anti-de\thinspace Sitter spacetime has been presented
  by Holst and Peld{\'a}n~\cite{hp:1997}; the CCBH manifold described by
  Ba{\~n}ados can be considered to be a submanifold of one of the solutions
  found in Ref.~\cite{hp:1997}.}

In this letter, we examine the thermodynamic properties of
the CCBH spacetime in General Relativity.
In general, a given black hole solution can arise from a variety of theories,
and its thermodynamic properties are theory-dependent. 
In order to understand the thermodynamic properties of CCBHs, 
Ba{\~n}ados considered the black hole to be a solution
of a five-dimensional Chern-Simons supergravity theory.  
In such a theory, the thermodynamic
variables can be constructed for a rotating solution, but the
result is surprising: the thermodynamic internal energy is associated with
the angular momentum parameter of the solution while the thermodynamic
conjugate to the angular velocity is associated with the mass parameter.
In addition, the entropy is found to be proportional to the circumference
of the \emph{inner} horizon rather than the outer horizon.
Such phenomena also occur for the 2+1 dimensional black hole when
the thermodynamic variables are computed from a Chern-Simons like
action~\cite{cgm:1995}, though a more conventional result for the
thermodynamic variables is obtained when the action of General Relativity
is used~\cite{bcm:1994}.  We consider here CCBHs in the context of
four-dimensional General Relativity, although our results may be
straightforwardly generalized to any larger number of dimensions.%
\footnote{The thermodynamics of other asymptotically anti-de\thinspace Sitter
  black holes with non-trivial spatial topology has been studied in
  Ref.~\cite{blp:1997}}

\begin{figure}
\begin{center}
\epsfig{file=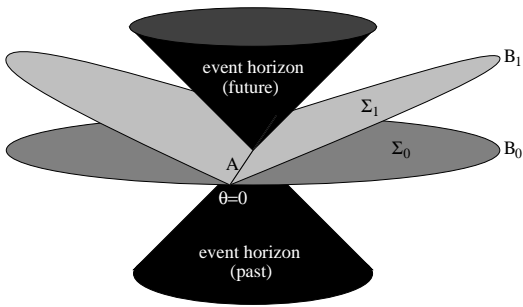,width=0.8\linewidth}
\end{center}
\caption{The Lorentzian CCBH spacetime.  The two cones represent the future and
  past event horizons of the black hole, while the spacelike
  surfaces $\Sigma_0$ and~$\Sigma_1$ are surfaces of constant coordinate
  time.  The singularity within the horizon is not shown.  Each point
  represents a circle in the suppressed coordinate~$\phi$.  The outer
  boundaries of $\Sigma_0$ and~$\Sigma_1$ are the quasilocal surfaces of
  constant time and radius.  The foliation is degenerate along the axis~$A$.}
\label{f:lorentz}
\end{figure}

For definiteness, we consider the non-rotating CCBH spacetime.
This spacetime has the line element
\begin{equation}
  ds^2 = \frac{\ell^4 f^2(r)}{\rH^2}[d\theta^2 - \sin^2\theta\,(dt/\ell)^2]
    + \frac{dr^2}{f^2(r)} + r^2 d\phi^2
  \label{e:metric}
\end{equation}
with the metric function
\begin{equation}
  f^2(r) = \frac{r^2 - \rH^2}{\ell^2}.
  \label{e:metricfn}
\end{equation}
The quantity $\rH$ is the circumferential radius of the ``bolt'' of the Killing
horizon, and $\ell$ is the length scale of the anti-de\thinspace Sitter
spacetime curvature.  The angle $\phi$ is periodic with period~$2\pi$; the
coordinate system is valid outside the black hole (i.e., for~$r>\rH$) and
for~$0<\theta<\pi$.  The details of the construction of this spacetime from
ordinary anti-de\thinspace Sitter spacetime can be found in Ref.~\cite{b:1997}.
Because this solution is merely anti-de\thinspace Sitter spacetime with
identifications, it is a solution to the field equations
arising from the Einstein-Hilbert action,
\begin{equation}
  I = \int_M \bbox{L}
    = \frac{1}{16\pi}\int_M {}^{4}\!\bbox{\epsilon}\,(R - 2\Lambda),
  \label{e:action}
\end{equation}
with cosmological constant~$\Lambda=-3/\ell^2$.
Here, $\bbox{L}$ is the Einstein-Hilbert Lagrangian 4-form (with a cosmological
constant) and ${}^{4}\!\bbox{\epsilon}$ is the volume form on the manifold~$M$.
The Lorentzian black hole spacetime is depicted in Fig.~\ref{f:lorentz}.
Notice that the foliation of the spacetime into leaves of constant coordinate
time~$t$ is degenerate on the axis~$A$ with~$\theta=0$ and~$\theta=\pi$ (we
have included a second side with $\pi<\theta<2\pi$ in Fig.~\ref{f:lorentz}).
The quasilocal surface~$B$ is taken to be a 2-surface of constant time and
radius~$r=R>\rH$.

\begin{figure}
\begin{center}
\epsfig{file=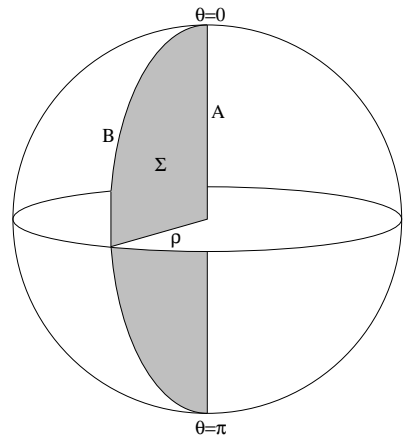,width=0.8\linewidth}
\end{center}
\caption{The Euclidean CCBH instanton.
  The azimuthal angle is the time~$\tau/\ell$, the polar angle is~$\theta$,
  and the radius is the proper radius~$\rho=\int dr/f(r)$
  with~$\rho(\rH)=0$.  Each point represents a circle in the suppressed
  coordinate~$\phi$.  The surface~$\Sigma$ is a surface of constant time;
  its boundary consists of the quasilocal surface~$B$ and the axis~$A$ of
  the sphere.}
\label{f:euclid}
\end{figure}

Let us begin our analysis of the properties of the CCBH with a calculation
of the entropy.  The entropy of a black hole spacetime is equal to the value
of the microcanonical
action of the Euclidean section of the spacetime~\cite{by:1993b}.  In the case
of the CCBH, the Euclidean section is obtained by the Wick rotation
$t\to\tau=it$.  The line element is
\begin{equation}
  ds^2 = \frac{\ell^4 f^2(r)}{\rH^2}[d\theta^2 + \sin^2\theta\,(d\tau/\ell)^2]
    + \frac{dr^2}{f^2(r)} + r^2 d\phi^2.
  \label{e:euclidmetric}
\end{equation}
Notice that the quantity in brackets is the line element of a two sphere if
$0\le\theta\le\pi$ and $\tau$ is periodic with period~$2\pi\ell$.  If such
an identification of the time is made, the Euclidean manifold is regular
and is depicted in Fig.~\ref{f:euclid}.%
\footnote{The surface gravity of the event horizon is
  $\kappa_{\scriptscriptstyle\mathrm{H}}=[-\frac{1}{2}
    (\nabla^at^b)(\nabla_at_b)]^{1/2}=1/\ell$, so the usual regularity
  condition $\Delta\tau=2\pi/\kappa_{\scriptscriptstyle\mathrm{H}}$ applies.}

The microcanonical action differs from the action of Eq.~(\ref{e:action})
by a boundary term on the history~$T$ of the
quasilocal surface~$B$ \cite{iw:1995}:
\begin{equation}
  I_{\text{micro}} = \int_M \bbox{L} - \int_T \bbox{d}t\wedge\bbox{q}[t].
  \label{e:microaction}
\end{equation}
The boundary functional contains the Noether charge 2-form, $\bbox{q}[t]$,
associated with the covariance of the Lagrangian under diffeomorphisms
generated by the vector~$t^a=(\partial/\partial t)^a$~\cite{w:1993}.  On a
two-dimensional submanifold with binormal~$n^{ab}$ and volume element
${}^2\!\epsilon_{ab}=\frac{1}{2}n^{cd}\,{}^4\!\epsilon_{cdab}$, the Noether
charge 2-form is given by
\begin{equation}
  \bbox{q}[t] = \frac{1}{16\pi}\,{}^2\!\bbox{\epsilon}\,n^{ab}\nabla_a t_b.
  \label{e:noethercharge}
\end{equation}
The microcanonical action can be evaluated on the Euclidean manifold to yield
the entropy.  We follow the method of Iyer and Wald~\cite{iw:1995} in computing
the entropy.  Because the spacetime is stationary, we find
\begin{equation}
  S = \Delta\tau \left[
    -\int_{\partial\Sigma} \bbox{q}[t] + \int_B \bbox{q}[t] \right].
  \label{e:entropyintegral}
\end{equation}
with $\Delta\tau=2\pi\ell$.
{}From Fig.~\ref{f:euclid}, it is clear that $\partial\Sigma$ contains two
pieces: the quasilocal surface~$B$ and the axis~$A$ of the spherical instanton.
Thus, the entropy only depends on the integral of the Noether charge 2-form
over the axis~$A$ of the spherical instanton.  The binormal to~$A$ is
$n^{ab}=2u^{[a}m^{b]}$ where $u^a$ is the normal vector to surfaces of constant
time, and $m^a$ is the normal vector to surfaces of constant~$\theta$.
The Noether charge 2-form is found to be
$\bbox{q}[t]={}^2\!\bbox{\epsilon}\,(8\pi\ell)^{-1}$ where
${}^2\!\bbox{\epsilon}$ is the area element of the 2-surface~$A$.  Integrating
the Noether charge over the boundary~$A$ (which consists of both the portion
with $\theta=0$ and~$\theta=\pi$), we find that the entropy is
\begin{equation}
  S = \pi\ell^2 f(R)
  \label{e:entropy}
\end{equation}
where $r=R$ is the radius of the quasilocal surface~$B$.  Notice that the
entropy depends on the size of the quasilocal surface: this dependence occurs
because the entropy is associated with the area of the cylinder~$A$ which
extends to~$r=R$.

We can also calculate the quasilocal thermodynamical variables in order to
verify that the first law of thermodynamics holds.  The relevant quantities
we need are the quasilocal energy density and the surface stress tensor.
These variables are calculated using the definitions of Brown and
York~\cite{by:1993a}.  The quasilocal energy density derived from the
Einstein-Hilbert action is given by
\begin{equation}
  {\mathcal{E}} = \frac{1}{8\pi}\sqrt{\sigma}\,k.
  \label{e:energydef}
\end{equation}
Here, $k$ is the trace of the extrinsic curvature~$k_{ab}$ of the quasilocal
surface~$B$ embedded in the spacelike surface~$\Sigma$:
$k_{ab}=-\sigma_a{}^cD_cn_b$ where $D_a$ is the covariant derivative operator
on~$\Sigma$, $n^a$ is the normal vector to $B$ embedded in~$\Sigma$, and
$\sigma_{ab}$ is the induced metric on~$B$.  Similarly, the quasilocal surface
stress tensor is
\begin{equation}
  {\mathcal{S}}^{ab} = \frac{1}{16\pi}\sqrt{\sigma}[k^{ab}
    - \sigma^{ab}(k-n^c a_c)]
  \label{e:stressdef}
\end{equation}
where $a_c=u^a\nabla_au_c$ is the acceleration of the timelike unit
normal~$u^a$ to the surfaces~$B$ embedded in~$T$.  In general, the quasilocal
energy density also has a contribution arising from an arbitrary background
action functional; this contribution effectively provides a zero point for the
energy in a reference spacetime.  However, it is difficult to choose a
reference spacetime for the CCBH because the
intrinsic geometry of the quasilocal surface~$B$ depends on the constant of
integration~$\rH$.  Fortunately, since the first law of thermodynamics only
depends on changes in the quasilocal energy, the contribution from the
reference spacetime is irrelevant when analyzing the thermodynamics of
the spacetime.

The calculation of the quasilocal energy density and surface stress tensor
is straightforward.  From Eqs.~(\ref{e:energydef}) and~(\ref{e:stressdef}),
we obtain
\begin{equation}
  {\mathcal{E}} = -\frac{1}{8\pi\rH}\,[R^2 + \ell^2 f^2(R)]
  \label{e:energy}
\end{equation}
for the quasilocal energy and
\begin{mathletters}
\label{e:stress}
\begin{eqnarray}
  {\mathcal{S}}^{\theta\theta} &=& \frac{1}{16\pi}\frac{\rH}{\ell^4 f^2(R)}\,
    [R^2 + \ell^2 f^2(R)]
  \label{e:stresstheta} \\
  {\mathcal{S}}^{\phi\phi} &=& \frac{1}{8\pi\rH}
  \label{e:stressphi}
\end{eqnarray}
\end{mathletters}%
for the quasilocal stress tensor.  In addition, the inverse
temperature~$\beta(R)$
on the quasilocal boundary can be computed: it is just the red-shifted period
of identification of the Euclidean time.  We find
\begin{equation}
  \beta(R) = [g_{tt}(R)]^{1/2}\Delta\tau = \frac{2\pi\ell^2 f(R)}{\rH}\,
    \sin\theta.
  \label{e:temperature}
\end{equation}
Notice that the temperature is not constant on the quasilocal surface.
In particular, it diverges at $\theta=0$ and~$\theta=\pi$.  This is because
the foliation becomes degenerate at these points.

The first law of thermodynamics is obtained by consideration of variations
of the microcanonical action evaluated on the Euclidean manifold.  As shown
in Refs.~\cite{by:1993b,iw:1995},
\begin{equation}
  \delta S = \int_0^\pi d\theta \int_0^{2\pi} d\phi\,\beta\,
    [\delta{\mathcal{E}}+{\mathcal{S}}^{ab}\delta\sigma_{ab}] .
  \label{e:firstlaw}
\end{equation}
Because the quasilocal boundary is not an isotherm, the first law of
thermodynamics must be left in an integral form, i.e., the temperature
cannot be factored out of the integral.
Eq.~(\ref{e:firstlaw}) can be explicitly verified using the quasilocal energy
density of Eq.~(\ref{e:energy}), the quasilocal stress tensor of
Eqs.~(\ref{e:stress}), the temperature of Eq.~(\ref{e:temperature}) and the
entropy of Eq.~(\ref{e:entropy}).  Recall that $\sigma_{ab}$ is the metric on
the quasilocal boundary~$B$.  The variations induced in the entropy, energy,
and metric~$\sigma_{ab}$ are variations in both the constant of
integration~$\rH$ and the size of the quasilocal system~$R$.  Two unusual
features of the CCBH spacetime thermodynamics are the
facts that the entropy depends on the size~$R$ of the quasilocal system and
that the metric of the quasilocal boundary depends on the constant of
integration~$\rH$.  Thus, under variations in the parameter~$\rH$ alone,
there is work done by the surface stress; similarly, a process involving a
change in the size of the quasilocal system alone is not adiabatic.

We have shown that it is possible to compute the thermodynamic variables
associated with the CCBH spacetime in 3+1 dimensions as a
solution to the theory of General Relativity.  In order to avoid the effects
of the unusual asymptotic behavior of the spacetime, we have adopted
quasilocal definitions of the thermodynamic variables.  When the spacetime
is foliated into leaves associated with the timelike Killing vector, the
Euclidean instanton has an unusual topology: the foliation becomes degenerate
on a cylinder that contains the bifurcation circle of the event horizon.
The entropy is associated with the area of this cylinder, and it vanishes
as the quasilocal surface approaches the horizon.  This result indicates
that the entropy is not so much associated with the black hole horizon
as with the area of the surface of degeneracy in the foliation of spacetime.
It should be emphasized that the foliation degeneracy results from the
requirement that the observers remain static; however, this requirement is
needed in order to treat the spacetime as a thermodynamic system.
In addition, the metric
on the quasilocal boundary depends on the constant of integration of the
black hole solution.  Because of this, it is difficult to find a reference
spacetime that produces a suitable zero-point for the quasilocal energy.
Nevertheless, the thermodynamic variables do satisfy the first law of
thermodynamics given in Eq.~(\ref{e:firstlaw}).

\acknowledgements
We would like to thank Peter Peld{\'a}n for his comments on an earlier
draft of this paper.
This work was supported in part by the Natural Sciences and Engineering
Research Council of Canada.  J.~Creighton gratefully acknowledges
partial support from NSF-grant AST-941731.


\begin{references}
\bibitem{b:1997} M. Ba{\~n}ados, Phys. Rev. D (to be published).
\bibitem{bhtz:1993} M. Ba{\~n}ados, M. Henneaux, C. Teitelboim, and J. Zanelli,
  Phys. Rev. D \textbf{48}, 1506 (1993).
\bibitem{hp:1997} S. Holst and P. Peld{\'a}n, Report No. gr-qc/9705067, 1997
  (unpublished).
\bibitem{cgm:1995} S. Carlip, J. Gegenberg, and R. Mann,
  Phys. Rev. D \textbf{51}, 6854 (1995).
\bibitem{bcm:1994} J. D. Brown, J. D. E. Creighton, and R. B. Mann,
  Phys. Rev. D \textbf{50}, 6394 (1994);  O. B. Zaslavskii, Class. Quantum
  Grav. \textbf{11}, L33 (1994).
\bibitem{blp:1997} D. Brill, J. Louko, and P. Peld{\'a}n,
  Phys. Rev. D \textbf{56}, 3600 (1997);
  L. Vanzo, Phys. Rev. D (to be published).
\bibitem{by:1993b} J. D. Brown and J. W. York, Phys. Rev. D \textbf{47} 1420
  (1993).
\bibitem{iw:1995} V. Iyer and R. M. Wald, Phys. Rev. D \textbf{52} 4430 (1995).
\bibitem{w:1993} R. M. Wald, Phys. Rev. D \textbf{48} 3427 (1993);
  V. Iyer and R. M. Wald, Phys. Rev. D \textbf{50} 846 (1994).
\bibitem{by:1993a} J. D. Brown and J. W. York, Phys. Rev. D \textbf{47} 1407
  (1993).
\end{references}
\end{document}